# Crystal Structure Manipulation of the Exchange Bias in an Antiferromagnetic Film

Wei Yuan [1,2], Tang Su[1,2], Qi Song[1,2], Wenyu Xing[1,2], Yangyang Chen[1,2], Tianyu Wang[1,2], Zhangyuan Zhang[3,4], Xiumei Ma[3], Peng Gao[2,3], Jing Shi[5*], and Wei Han[1,2*]

[1]International Center for Quantum Materials, Peking University, Beijing, 100871, P. R. China

[2]Collaborative Innovation Center of Quantum Matter, Beijing 100871, P. R. China

[3]Electron Microscopy Laboratory, School of Physics, Peking University, Beijing, 100871, P. R. China

[4]Department of Physics and Key Laboratory of Artificial Micro- and Nano-structures of Ministry of Education, Wuhan University, Wuhan 430072, P. R. China

[5]Department of Physics and Astronomy, University of California, Riverside, California 92521, USA

*Correspondence to be addressed to: jing.shi@ucr.edu (J.S.) and weihan@pku.edu.cn (W.H.)

**Abstract**

Exchange bias is one of the most extensively studied phenomena in magnetism, since it exerts a unidirectional anisotropy to a ferromagnet (FM) when coupled to an antiferromagnet (AFM) and the control of the exchange bias is therefore very important for technological applications, such as magnetic random access memory and giant magnetoresistance sensors. In this letter, we report the crystal structure manipulation of the exchange bias in epitaxial hcp $Cr_2O_3$ films. By epitaxially growing twined ($10\bar{1}0$) oriented $Cr_2O_3$ thin films, of which the $c$ axis and spins of the Cr atoms lie in the film plane, we demonstrate that the exchange bias between $Cr_2O_3$ and an adjacent permalloy layer is tuned to in-plane from out-of-plane that has been observed in (0001) oriented $Cr_2O_3$ films. This is owing to the collinear exchange coupling between the spins of the Cr atoms and the adjacent FM layer. Such a highly anisotropic exchange bias phenomenon is not possible in polycrystalline films.

**Introduction**

Exchange bias refers to the shift of the magnetic hysteresis loop of a ferromagnetic (FM) layer away from the zero magnetic field, resulting from the exchange interaction from an antiferromagnet (AFM) layer[1-4]. In practical applications such as magnetic random access memory and giant magnetoresistance sensors, etc.[5,6], the exchange bias is used to "pin" the FM magnetization from switching in small magnetic fields so that the FM layer could serve as a fixed reference layer. In previous studies, it has been established that the exchange bias hinges on the spin orientations of the surface magnetic atoms in the AFM layer[7-13]. Since the direction of the spin orientations of the surface magnetic atoms is associated with the crystal structure of the AFM layer, the highly anisotropic exchange bias could be simply manipulated by the crystal orientation design. An ideal candidate AFM material to achieve this objective is the single crystalline $Cr_2O_3$ films with a hexagonal close packed (hcp) structure, of which the spin orientation of the Cr atoms is uniquely directed along the $c$ axis and could dictate the direction of the exchange bias[7,14,15].

In this letter, we report the manipulation of the exchange bias by controlling the surface spin configuration via crystal orientation design. By epitaxially growing $(10\bar{1}0)$ oriented $Cr_2O_3$ films on (001) oriented rutile $TiO_2$ substrates, we force the $c$ axis to lie in the film plane, which results in a strong in-plane exchange bias between the $Cr_2O_3$ film and an adjacent permalloy (Py) layer, whilst the perpendicular exchange bias is completely suppressed. The perpendicular exchange bias was previously shown in (0001) oriented $Cr_2O_3$ films[1-4,7,15,16]. Our results along with the previous studies demonstrate crystal structure manipulations of the exchange bias based on the collinear exchange coupling between the spins of the Cr atoms and the adjacent FM layer.

**Results and Discussion**

The ($10\bar{1}0$) oriented $Cr_2O_3$ films are grown on the (001) oriented rutile $TiO_2$ substrates via laser molecular beam epitaxy (LMBE) with a base pressure of $2 \times 10^{-8}$ mbar (see methods for details). Fig. 1a and 1b show the *in-situ* reflection high-energy electron diffraction (RHEED) characterization of the (001) oriented $TiO_2$ substrate's surface viewed from the [100] and [110] directions. After the initial growth of 3 nm $Cr_2O_3$, the RHEED pattern of the $TiO_2$ substrate disappears and that of $Cr_2O_3$ starts to appear, as shown in Fig. 1c and 1d. As the thickness of $Cr_2O_3$ film increases, its RHEED pattern becomes brighter. Fig. 1e-1h show the RHEED patterns of 10 nm and 27 nm $Cr_2O_3$ films, respectively. To be noted, four satellite RHEED spots are observed around each main diffraction spot viewed from the [100] direction of the $TiO_2$ substrates (Fig. 1e and 1g).

The crystalline structural properties of these $Cr_2O_3$ films are further characterized by x-ray diffraction (XRD). The θ - 2θ scans of the rutile $TiO_2$ substrate, 10 nm, 20 nm, and 27 nm $Cr_2O_3$ films are shown in Fig. 2a. The peak at 2θ of ~ 63 degrees corresponds to the (002) peak of the $TiO_2$ substrates. For the 10 nm $Cr_2O_3$ film, a peak at 2θ of ~ 65 degrees is observed, which corresponds to the ($30\bar{3}0$) peak of the $Cr_2O_3$ crystal. As the thickness of $Cr_2O_3$ increases, the intensity of the peak at ~ 65 degrees becomes stronger. We note that for the 27 nm $Cr_2O_3$ film, only ($30\bar{3}0$) peak is detectable for the whole scan range (see supplementary information; Fig. S1), indicating good crystalline properties of the ($10\bar{1}0$) oriented $Cr_2O_3$ thin films. The surface morphology is characterized by atomic force microscopy (AFM). The root-mean-square (RMS) roughness is 0.09 nm for the rutile $TiO_2$ substrate after annealing in the chamber (Fig. 2b). After the growth of 13 nm $Cr_2O_3$ film, the RMS roughness increases to 0.28 nm (Fig. 2c), indicating that the surface of the epitaxial $Cr_2O_3$ films is quite smooth.

The epitaxial growth of the $(10\bar{1}0)$ oriented $Cr_2O_3$ films on $TiO_2$ is quite interesting, given the fact that $Cr_2O_3$ and $TiO_2$ belong to totally different space groups. $Cr_2O_3$ has a hexagonal crystal structure, which belongs to the $R\bar{3}c$ group[17], while rutile $TiO_2$ has a cubic structure, which belongs to the $P4_2$ group[18]. However, the $c$ lattice constant of $Cr_2O_3$ is 13.599 Å, and the $a$ lattice constant of $TiO_2$ is 4.584 Å, which results in a coincidental anion alignment with a lattice mismatch of only ~1.1% ($\frac{|13.599-3\times 4.584|}{13.599} \times 100\%$). Hence, the $Cr_2O_3$ films could be grown with the $c$ axis lying in-plane and parallel to $a$ or $b$ axis of the $TiO_2$ substrates (Fig. 3a). As the $TiO_2$ crystal's $ab$ plane has four-fold symmetry, which could result in four-fold in-plane rotational symmetry of the crystalline structure of the $Cr_2O_3$ thin films. To investigate this, HRTEM is used to characterize the interfacial structure properties between $Cr_2O_3$ and $TiO_2$ viewed from the [010] direction of the $TiO_2$ substrate. As shown in Fig. 3b, a sharp interface with $TiO_2$ is observed, as indicated by the pink dashed line. For $Cr_2O_3$, two distinct zones of the crystalline structures are observed, one of which is denoted as zone [0001], and the other one is denoted as zone $[1\bar{2}10]$. The boundary of these two zones is marked by a yellow dashed line. For zone [0001], the $c$ axis of $Cr_2O_3$ is parallel to the [010] direction of the $TiO_2$ substrate, and the six-fold symmetry pattern of the basal plane can be identified. For zone $[1\bar{2}10]$, the $c$ axis is parallel to the [100] direction of the $TiO_2$ substrate. The crystal orientation of $Cr_2O_3$ films has been shown to be highly associated to the substrate crystalline structures. For example, preferential (0001) oriented growth of $Cr_2O_3$ is achieved on (0001) oriented $Al_2O_3$, Co, and (111) oriented Cu and $(11\bar{2}0)$ oriented $Cr_2O_3$ is grown on Fe (110) films[15,19,20]. The growth mode of $(10\bar{1}0)$ $Cr_2O_3$ film in our study is similar to that observed in an earlier report for the ($10\bar{1}0$) oriented $Fe_2O_3$ films grown on rutile (001) $TiO_2$ substrates[21].

Interestingly, with the *c* axis lying in the film plane, the spin orientations of the Cr atoms also lie in-plane in these ($10\bar{1}0$) oriented $Cr_2O_3$ films, as schematically shown in Fig. 3a. This is very different from previously reported (0001) oriented $Cr_2O_3$ films grown on $Al_2O_3$ substrates, of which both the spin orientations and the exchanges bias are perpendicular to the films[7,15,16]. To study the exchange bias of the epitaxial $Cr_2O_3$ films (see methods for details), we deposit 10 nm Py on top of the ($10\bar{1}0$) oriented $Cr_2O_3$ films and measure the magnetic hysteresis loops by Magnetic Properties Measurement System (MPMS; Quantum Design) with both in-plane and out-of-plane magnetic fields at various temperatures (schematic drawings shown in Fig. 4a and 4b). The in-plane magnetic hysteresis loop is first measured. Prior to the measurement, the sample is cooled from 400 to 10 K in an in-plane magnetic field of 1000 Oe along the [100] direction of the $TiO_2$ substrate, which is much smaller than the spin-flop field of several Tesla for $Cr_2O_3$ reported previously[22]. By cooling through the blocking temperature, the magnetization direction of the Py sets the surface spin configurations of the $Cr_2O_3$ films. Then, we measure the magnetization of the Py as a function of the in-plane magnetic field along the [100] direction of the $TiO_2$ substrate (Fig. 4a) from 10 to 300 K. After subtracting a linear background which is mainly due to the diamagnetic response of the rutile $TiO_2$ substrate, the shifted magnetic hysteresis loops of Py are displayed in Fig. 4c. At 10 K, as the magnetic field ramps from negative to positive, a sharp jump in magnetic moment occurs at ~110 Oe, but the jump occurs at ~ -400 Oe on the return sweep. These two switching fields are labeled as $H_1$ and $H_2$, respectively, as indicated in the top panel of the Fig. 4c. The exchange bias field ($H_B$) is defined by the mean value of the $H_1$ and $H_2$, i.e. $H_B = (H_1 + H_2)/2$. As the temperature increases, the exchange bias field steadily decreases from the low temperature value.

To characterize the anisotropy of the exchange bias effect, magnetic hysteresis loops are measured with a magnetic field perpendicular to the films (Fig. 4b). The same field cooling procedure as with the in-plane magnetic fields is adopted. The out-of-plane magnetization curves measured at 10, 20, 40, and 60 K are shown in Fig. 4d. The symmetric magnetization hysteresis loops indicate a negligible perpendicular exchange, which is in stark contrast to the out-of-plane exchange bias observed in the (0001) oriented $Cr_2O_3$ films. The highly anisotropic exchange bias phenomenon can be attributed to the crystalline orientation difference of the $Cr_2O_3$ films. In hcp structures, the c-axis direction dictates the spin orientation of the magnetic atoms. In the (0001) oriented $Cr_2O_3$ films, the spin orientation of the Cr atoms is perpendicular to the films, whereas in the ($10\bar{1}0$) oriented $Cr_2O_3$ films, the spins of the Cr atoms lie in the film plane. These results further indicate the direct collinear exchange coupling between the spins of the Cr atoms and the adjacent Py layer.

Fig. 5a and 5b summarize the in-plane exchange bias field and in-plane conceive field ($H_C$), where $H_C = (|H_1 - H_2|)/2$, for the sample consisting of the 13 nm $Cr_2O_3$ films and 10 nm Py as a function of the temperature. As the temperature increases from 10 to 60 K, the in-plane exchange bias field (Fig. 5a, Blue dots) decreases quickly from ~ -150 Oe to almost 0 Oe, where the sign depends on the magnetic field direction during magnetic cooling. No exchange bias is observable at and above 60 K, which implies a blocking temperature ($T_B$) of ~ 60 K. An abrupt increase in $Hc$ below 60 K is another property of exchange biased Py, which is due to the formation of the AFM order in this 13 nm $Cr_2O_3$ thin film[2]. As there are two crystalline zones of ($10\bar{1}0$) oriented $Cr_2O_3$, as indicated in zones [0001] and [$1\bar{2}10$], we also measure the exchange bias in the direction along the $TiO_2$ [010] direction. Almost identical exchange biases are observed at each temperature (Fig. 5a, Green dots).

The measured $T_B$ of 13 nm Cr$_2$O$_3$ is ~60 K, which is much lower compared to the value reported on (0001) oriented bulk Cr$_2$O$_3$ single crystals[15]. In antiferromagnetic films, it has been known that $T_B$ is highly related to Neel temperature ($T_N$), and is usually slightly lower than the $T_N$. Both $T_B$ and $T_N$ increase as the AFM thickness increases due to finite-size effects[2,12,23,24]. To obtain the $T_B$ as a function of the thicknesses of the Cr$_2$O$_3$ thin films, the in-plane exchange bias for the samples consisting of 7, 10, 20, and 27 nm Cr$_2$O$_3$ films and 10 nm Py bilayer films are also measured. Fig. 6a and 6b show the exchange bias as a function of temperature for the bilayer structures consisting of Cr$_2$O$_3$ (7 nm)/ Py(10 nm) and Cr$_2$O$_3$ (27 nm)/ Py(10 nm), respectively. The blocking temperatures of these two structures are determined to be 40 K and 100 K. The blocking temperature increases as the thickness of the Cr$_2$O$_3$ films increases, as shown in Fig. 6c. For the 27 nm Cr$_2$O$_3$ film, the blocking temperature is only ~ 100 K, which is far below the T$_B$ of bulk Cr$_2$O$_3$. One possible reason is the non-trivial finite size effects arising from the grain boundaries or oxygen defects in the Cr$_2$O$_3$[25].

**Conclusion**

In summary, we have demonstrated the manipulation of the exchange bias of the Cr$_2$O$_3$ thin films by controlling the surface spin orientation of the Cr atoms via crystal orientation design. For the epitaxial growth of ($10\bar{1}0$) oriented Cr$_2$O$_3$ films, the spin configurations of Cr atoms give rise to only in-plane exchange bias at the interface between Py and the Cr$_2$O$_3$ thin films, while no perpendicular exchange bias is observed. Our results along with previous studies on (0001) oriented Cr$_2$O$_3$ films indicate the collinear exchange coupling between the spins of the Cr atoms and the adjacent FM layer.

## Methods

**Cr$_2$O$_3$ films growth.** The (10$\bar{1}$0) oriented Cr$_2$O$_3$ films are grown on the (001) oriented rutile TiO$_2$ substrates via laser molecular beam epitaxy (LMBE) with a base pressure of $2 \times 10^{-8}$ mbar. Prior to the Cr$_2$O$_3$ growth, the substrate temperature is increased to 350 °C with a rate of 20 °C/min in the chamber with an oxygen partial pressure of 0.08 mbar. Then the Cr$_2$O$_3$ film is deposited from a Cr$_2$O$_3$ target with a laser power of $(8.0 \pm 0.2)$ mJ and a frequency of 2.0 Hz. The thickness of the Cr$_2$O$_3$ thin film ($t$) is determined from the cross section high resolution transmission electron microscopy.

**Exchange bias measurement.** A 10 nm Py is grown on top of the (10$\bar{1}$0) oriented Cr$_2$O$_3$ films by RF magnetron sputtering with a growth rate of 0.02 Å/s. A capping layer of 20 nm aluminum is deposited prior to taking the samples out of this sputtering chamber to prevent oxidization of Py. Magnetic Properties Measurement System (MPMS; Quantum Design) is used to measure the magnetic hysteresis loops to determine the exchange bias.

## Contributions

J.S. and W.H. proposed and supervised the studies. W. Y. grew the Cr$_2$O$_3$ films. T.S. and Q. S. grew the Py films. W.Y., T.S., Q.S., W.X, Y.C., and T.W. performed the exchange bias measurement and analyzed the data. Z.Z., X.M., and P.G. did the HRTEM measurement. W.Y., J.S. and W.H. wrote the manuscript.

## Acknowledgements

We acknowledge the funding support of National Basic Research Programs of China (973 Grants 2013CB921903, 2015CB921104, and 2014CB920902) and the National Natural Science

Foundation of China (NSFC Grant 11574006). Wei Han also acknowledges the support by the 1000 Talents Program for Young Scientists of China.

**Competing financial interests**

The authors declare no competing financial interests.

**References:**

1. Meiklejohn, W. H. & Bean, C. P. New Magnetic Anisotropy. *Phys. Rev.* **105**, 904-913, (1957).
2. Nogués, J. & Schuller, I. K. Exchange bias. *J. Magn. Magn. Mater.* **192**, 203-232, (1999).
3. Kiwi, M. Exchange bias theory. *J. Magn. Magn. Mater.* **234**, 584-595, (2001).
4. Radu, F. & Zabel, H. *Exchange Bias Effect of Ferro-/Antiferromagnetic Heterostructures*. 97-184 (Springer Berlin Heidelberg, 2008).
5. Parkin, S. S. P. *et al.* Exchange-biased magnetic tunnel junctions and application to nonvolatile magnetic random access memory (invited). *J. Appl. Phys.* **85**, 5828-5833, (1999).
6. Wolf, S. A. *et al.* Spintronics: A Spin-Based Electronics Vision for the Future. *Science* **294**, 1488, (2001).
7. Shiratsuchi, Y., Nakatani, T., Kawahara, S.-i. & Nakatani, R. Magnetic coupling at interface of ultrathin Co film and antiferromagnetic $Cr_2O_3$(0001) film. *J. Appl. Phys.* **106**, 033903, (2009).
8. Nogués, J., Lederman, D., Moran, T. J., Schuller, I. K. & Rao, K. V. Large exchange bias and its connection to interface structure in $FeF_2$–Fe bilayers. *Appl. Phys. Lett.* **68**, 3186-3188, (1996).
9. van der Zaag, P. J., Ball, A. R., Feiner, L. F., Wolf, R. M. & van der Heijden, P. A. A. Exchange biasing in MBE grown $Fe_3O_4$/CoO bilayers: The antiferromagnetic layer thickness dependence. *J. Appl. Phys.* **79**, 5103-5105, (1996).
10. Berkowitz, A. E. & Takano, K. Exchange anisotropy — a review. *J. Magn. Magn. Mater.* **200**, 552-570, (1999).
11. Morales, R. *et al.* Exchange-Bias Phenomenon: The Role of the Ferromagnetic Spin Structure. *Phys. Rev. Lett.* **114**, 097202, (2015).
12. Wu, J. *et al.* Direct Measurement of Rotatable and Frozen CoO Spins in Exchange Bias System of CoO/Fe/Ag (001). *Phys. Rev. Lett.* **104**, 217204, (2010).
13. Nogués, J., Moran, T. J., Lederman, D., Schuller, I. K. & Rao, K. V. Role of interfacial structure on exchange-biased $FeF_2$-Fe. *Phys. Rev. B* **59**, 6984-6993, (1999).
14. McGuire, T. R., Scott, E. J. & Grannis, F. H. Antiferromagnetism in a $Cr_2O_3$ Crystal. *Phys. Rev.* **102**, 1000-1003, (1956).

15  He, X. *et al.* Robust isothermal electric control of exchange bias at room temperature. *Nat. Mater.* **9**, 579-585, (2010).
16  Shiratsuchi, Y. *et al.* Detection and *In Situ* Switching of Unreversed Interfacial Antiferromagnetic Spins in a Perpendicular-Exchange-Biased System. *Phys. Rev. Lett.* **109**, 077202, (2012).
17  Newnham E, E. & Haan Y. M, D. E. in *Zeitschrift für Kristallographie - Crystalline Materials* Vol. 117   235 (1962).
18  Swope, R. J., Smyth, J. R. & Larson, A. C. H in rutile-type compounds; I, Single-crystal neutron and X-ray diffraction study of H in rutile. *American Mineralogist* **80**, 448-453, (1995).
19  Chen, X. *et al.* Ultrathin chromia films grown with preferential texture on metallic, semimetallic and insulating substrates. *Mater. Chem. Phys.* **149–150**, 113-123, (2015).
20  Sahoo, S., Mukherjee, T., Belashchenko, K. D. & Binek, C. Isothermal low-field tuning of exchange bias in epitaxial Fe∕Cr2O3∕Fe. *Appl. Phys. Lett.* **91**, 172506, (2007).
21  Williams, J. R., Wang, C. & Chambers, S. A. Heteroepitaxial growth and structural analysis of epitaxial α–Fe2O3(1010) on TiO2(001). *J. Mater. Res.* **20**, 1250-1256, (2005).
22  Seki, S. *et al.* Thermal generation of spin current in an antiferromagnet. *Phys. Rev. Lett.* **115**, 266601, (2015).
23  Ambrose, T. & Chien, C. L. Finite-Size Effects and Uncompensated Magnetization in Thin Antiferromagnetic CoO Layers. *Phys. Rev. Lett.* **76**, 1743-1746, (1996).
24  Imakita, K.-i., Tsunoda, M. & Takahashi, M. Thickness dependence of exchange anisotropy of polycrystalline Mn3Ir/Co-Fe bilayers. *J. Appl. Phys.* **97**, 10K106, (2005).
25  He, X., Echtenkamp, W. & Binek, C. Scaling of the Magnetoelectric Effect in Chromia Thin Films. *Ferroelectrics* **426**, 81-89, (2012).

**Figure Captions:**

**Fig. 1.** *In-situ* **RHEED characterization for the ($10\bar{1}0$) oriented Cr$_2$O$_3$ films grown on (001) oriented rutile TiO$_2$ substrates. a-b**, The RHEED patterns of the rutile TiO$_2$ substrate viewed from [100] and [110] directions prior to the Cr$_2$O$_3$ growth. **c-h**, The RHEED patterns of 3 nm (**c, d**), 10 nm (**e, f**), and 27 nm (**g, h**) Cr$_2$O$_3$ films, respectively. The figures in the left/right column are RHEED patterns viewed from TiO$_2$ [100]/[110] direction.

**Fig. 2. The crystalline structure and surface roughness of the ($10\bar{1}0$) oriented Cr$_2$O$_3$ films. a,** X-ray diffraction measurement of the rutile TiO$_2$ substrate and the 10 nm, 20 nm, and 27 nm Cr$_2$O$_3$ films, respectively. **b-c**, AFM images of a typical TiO$_2$ substrate and a 13 nm Cr$_2$O$_3$ film.

**Fig. 3. Crystalline structure of the ($10\bar{1}0$) oriented Cr$_2$O$_3$ film by HRTEM. a,** The schematic drawing showing the atomic interface between Cr$_2$O$_3$ and TiO$_2$. The spins of the Cr are parallel to the film plane, indicated by the red arrow. **b,** HRTEM characterization. The pink dashed line indicates the interface between Cr$_2$O$_3$ and TiO$_2$, and the yellow dashed line indicates the crystalline boundary between the zones [$1\bar{2}10$] and [0001] for the ($10\bar{1}0$) oriented Cr$_2$O$_3$ film. In the zone [$1\bar{2}10$], the [0001] directions of Cr$_2$O$_3$ is parallel to the [010], or [$0\bar{1}0$] of the TiO$_2$ substrate. Whilst in the zone [0001], the [0001] direction of Cr$_2$O$_3$ is parallel to the [100], or [$\bar{1}00$] of the TiO$_2$ substrate.

**Fig. 4. The characterization of the exchange bias for the sample of TiO$_2$/Cr$_2$O$_3$ (13 nm)/Py (10 nm)/ Al (20 nm). a-b,** Schematic drawing of the sample structure and the measurement geometry for in-plane exchange bias and perpendicular exchange bias, respectively. **c,** The magnetization hysteresis loops measured as a function of the in-plane magnetic field along the TiO$_2$ [100] direction at 10K, 20K, 40K, and 60 K, respectively. $H_1$ and $H_2$ indicate the coercive fields for the magnetization of Py and $H_B$ indicates the exchange bias. **d,** The magnetization curves measured as a function of the out-of-plane magnetic at 10K, 20K, 40K, and 60 K, respectively.

**Fig. 5 The exchange bias and the coercive field as a function of temperature for the sample TiO$_2$/Cr$_2$O$_3$ (13 nm)/ Py (10 nm)/ Al (20 nm). a,** The exchange bias field as a function of the temperature for magnetic field along the TiO$_2$ [100], and [010] directions, respectively. $T_B$ indicates the blocking temperature, above which the exchange bias becomes zero. **b,** The coercive field of the Py as a function of the temperature.

**Fig. 6. The exchange bias and blocking temperatures for the Cr$_2$O$_3$ films of various thicknesses. a-b,** The exchange bias as a function of the temperature for 7 nm and 27 nm Cr$_2$O$_3$ films, respectively. **c,** The blocking temperature as a function of the Cr$_2$O$_3$ film thicknesses ($t$) for the samples TiO$_2$/ Cr$_2$O$_3$ ($t$)/ Py (10 nm)/ Al (20 nm).

Figure 1

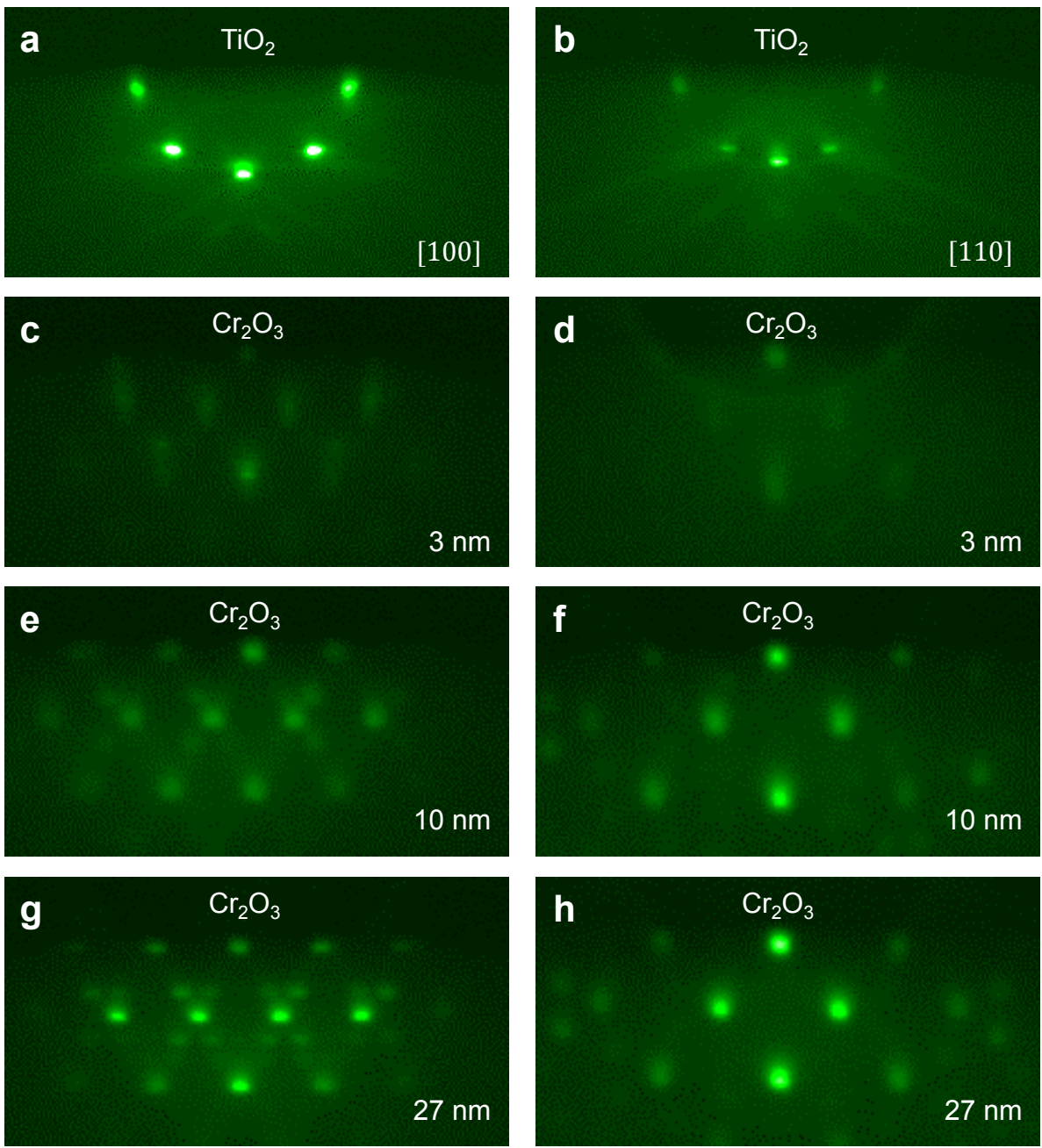

Figure 2

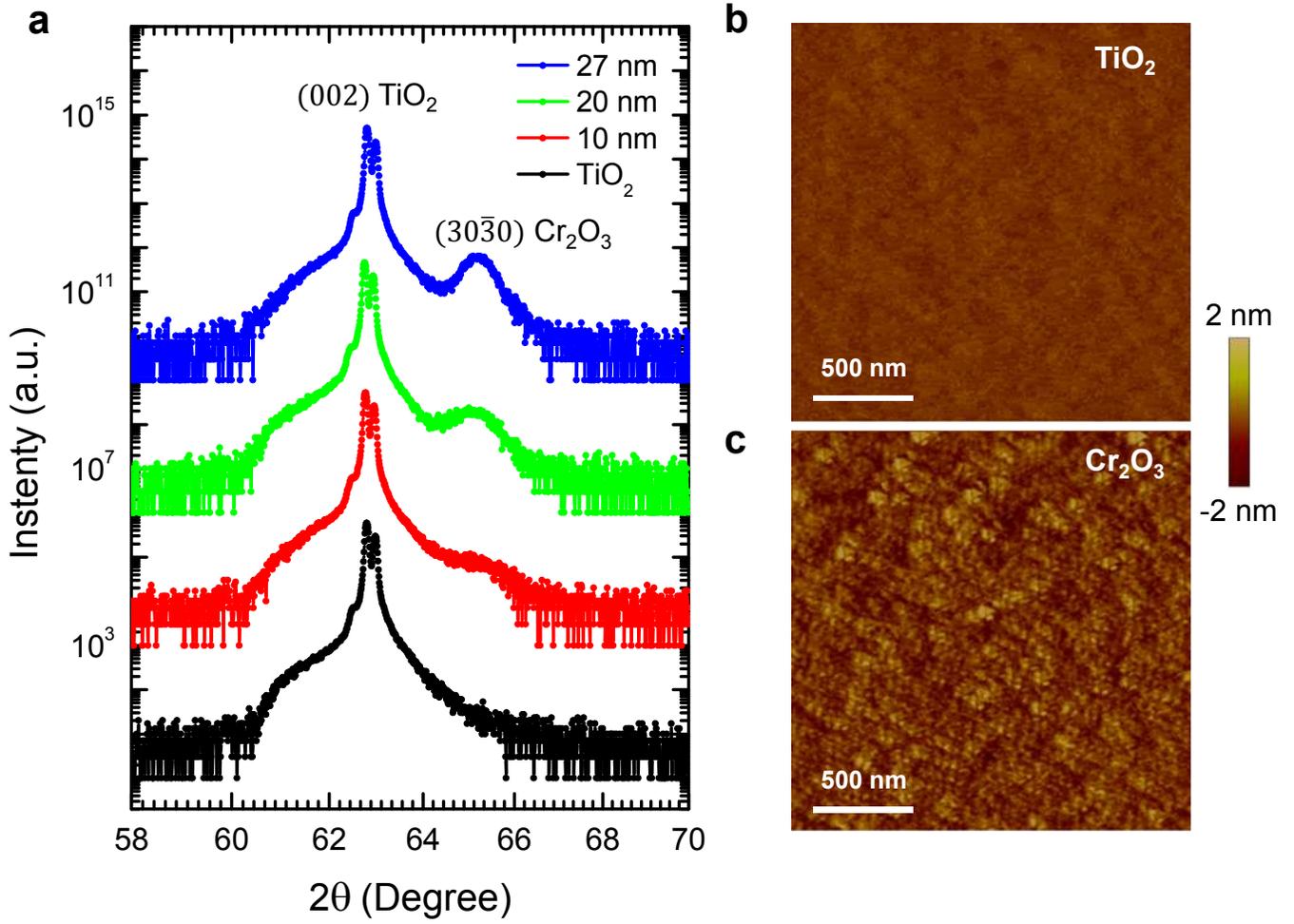

Figure 3

**a**

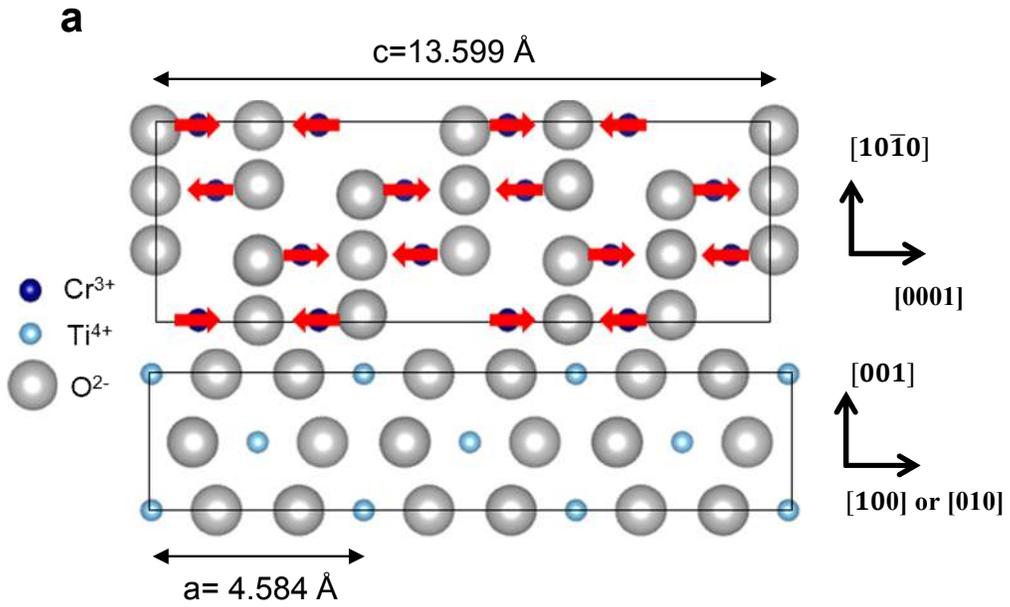

c=13.599 Å

- Cr³⁺ (Cr$^{3+}$)
- Ti⁴⁺ (Ti$^{4+}$)
- O²⁻ (O$^{2-}$)

a= 4.584 Å

$[10\bar{1}0]$
$[0001]$

$[001]$
$[100]$ or $[010]$

**b**

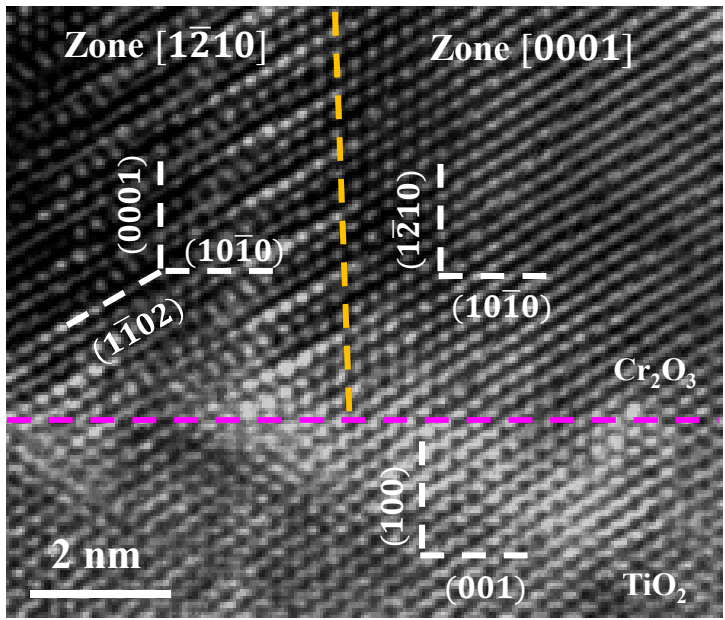

Zone $[1\bar{2}10]$  Zone $[0001]$

$(0001)$  $(1\bar{2}10)$
$(10\bar{1}0)$  $(10\bar{1}0)$
$(1\bar{1}02)$

$Cr_2O_3$

$(100)$
$(001)$  $TiO_2$

2 nm

Figure 4

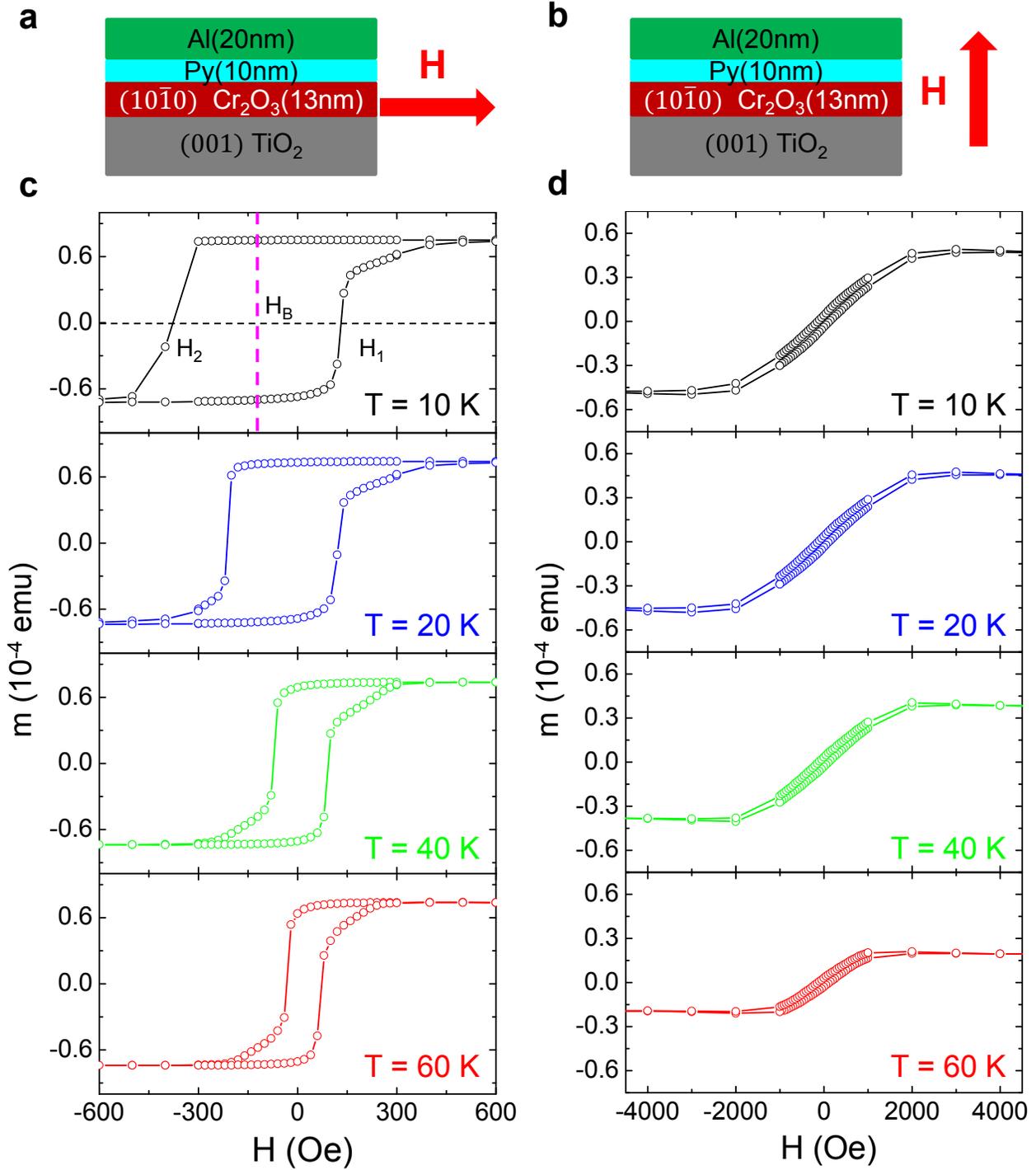

Figure 5

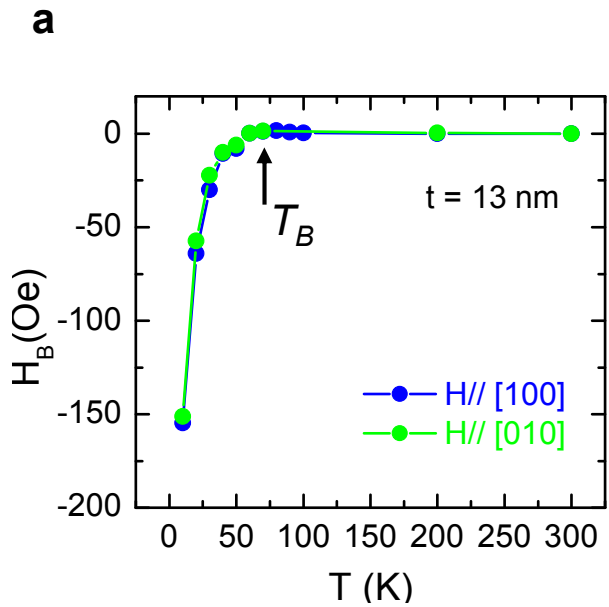

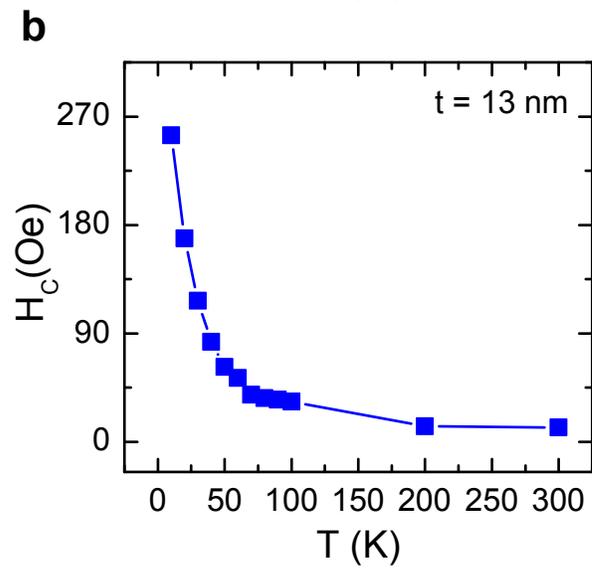

Figure 6

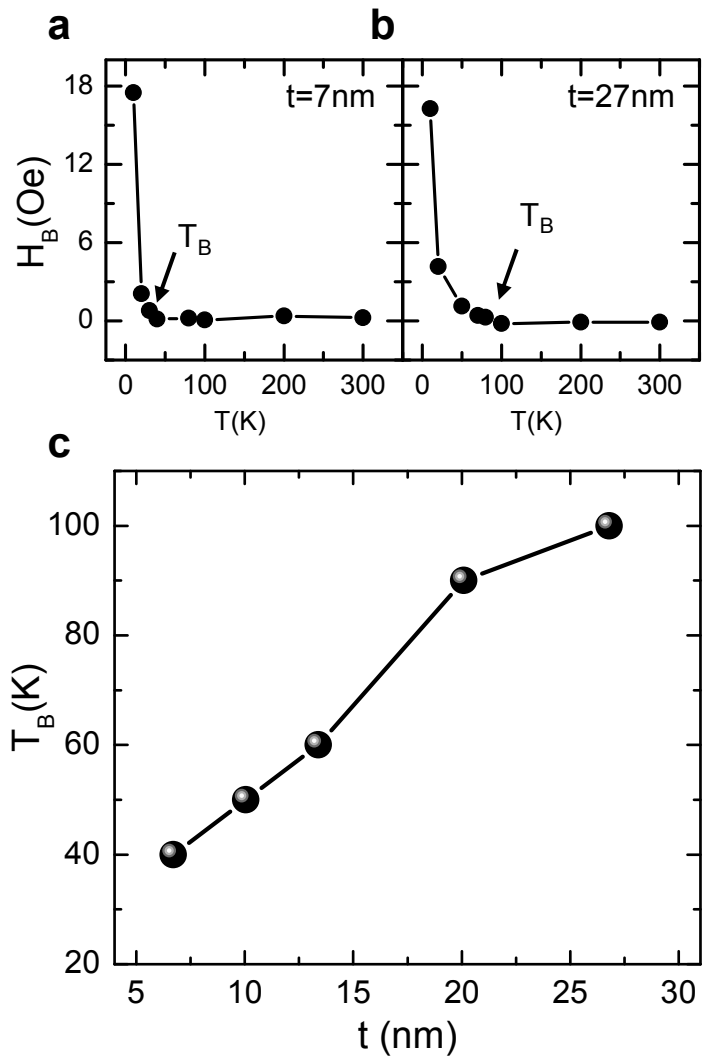